\documentclass[10pt]{iopart}

\pdfoutput=1

\usepackage{graphicx}  
\usepackage{dblfloatfix}

\newcommand{\E}{$E_\mathrm{app}$}

\newcommand{\erfc}{\mathrm{Erfc}}

\begin{document}

\title[Adsorbate dynamics on a silica-coated gold surface]
{Adsorbate dynamics on a silica-coated gold surface measured by Rydberg Stark spectroscopy }

\author{J. Naber, S. Machluf, L. Torralbo-Campo\footnote{Present address: CQ Center for Collective Quantum Phenomena and Their Applications, Physikalisches Institut, Eberhard-Karls-Universit{\"a}t T{\"u}bingen, Auf der Morgenstelle 14, D-72076 T{\"u}bingen, Germany},  M. L. Soudijn, N. J. van Druten, H. B. van Linden van den Heuvell, and R. J. C. Spreeuw}

\address{Van der Waals-Zeeman Institute, University of Amsterdam, Science Park 904, PO Box 94485,
1090 GL Amsterdam, The Netherlands}
\ead{r.j.c.spreeuw@uva.nl}
\vspace{10pt}
\begin{indented}
\item[]\today
\end{indented}

\begin{abstract}
Trapping a Rydberg atom close to a surface is an important step towards the realisation of many proposals of quantum information or hybrid quantum systems. 
One of the challenges in these experiments is to overcome the electric field emanating from contaminations on the surface. 
Here we report on measurements of an electric field created by $^{87}$Rb atoms absorbed on a 25$\,$nm thick layer of SiO$_2$, covering a 90$\,$nm layer of Au. 
The electric field is measured using a two-photon transition to the 23$D_{5/2}$ and 25$S_{1/2}$ state.
The electric field value that we measure is higher than typical values measured above metal surfaces, but is consistent with other measurements above SiO$_2$ surfaces.
In addition, we measure the temporal behaviour of the field and observe that we can reduce it in a single experimental cycle, using UV light or by mildly heating the surface, whereas the buildup of the field takes thousands of cycles.
We explain these results by a  change in the ad-atoms distribution on the surface.
These results indicate that the stray electric field can be reduced, opening new possibilities for experiments with trapped Rydberg atoms near surfaces.
\end{abstract}

\ioptwocol

\section{Introduction}

The investigation of Rydberg atoms close to a surface is  of great importance and interest for areas ranging from surface physics \cite{Hogan:2012iq,Pu:2013hf,So:2011bc} to quantum information \cite{Saffman:2010ky,Chan:2014kb,HermannAvigliano:2014bc,Thiele:2014kf,Thiele:2015ub}, particularly  in the context of atom chip \cite{Folman:2002tx,Reichel:2002kn} experiments. 
Their strong interaction over large distances \cite{Gallagher:2009jj} makes them a natural candidate for  efficient entanglement mechanisms in cold atom physics \cite{Jaksch:2000eg,Lukin:2001bu,Urban:2009jd,Gaetan:2009bla}. 
They can also help to investigate the transport of excitation energy \cite{Gunter:2013fv}. 
In addition, their high sensitivity to environmental influences \cite{Gallagher:2009jj} makes Rydberg atoms suitable as a surface probe.

However, noise sources can strongly influence Rydberg atoms' energy levels.
A typical example of a noise source in atom chip experiments is stray electric fields caused by ad-atoms that stick to the surface during the experimental cycle \cite{McGuirk:2004dm,Obrecht:2007br}. 
The interaction between the ad-atoms and the surface leads to effective electric dipoles on the surface, building up to a macroscopic electric field in the proximity of the surface \cite{Tauschinsky:2010ep,Hattermann:2012ho}. 
Atom chip experiments have to address this issue, and several methods to reduce the surface fields have been suggested \cite{Obrecht:2007br,HermannAvigliano:2014bc,Sedlacek:2015tg}. 

Here we report on the measurement of electric stray fields emanating from a surface, which consists of a 90$\,$nm Au layer covered with 25$\,$nm of SiO$_2$, see figure \ref{fig:ODimage}(a). 
We measure the Stark shift, induced by DC electric fields, of the 23$D_{5/2}$ and 25$S_{1/2}$ states of $^{87}$Rb and retrieve field gradients perpendicular and parallel to the atom chip. From our measurements we infer that the Rb ad-atoms are the main source for the stray electric fields. 
Then we examine several methods for reducing the stray fields. We observe the influence of ultra-violet (UV) light at 365$\,$nm, provided by an array of LEDs, and of the 480$\,$nm excitation laser on the stray fields. We put a special emphasis on the temporal dynamics of both effects.

\section{Methods}
\begin{figure*}[t]
\centering
\includegraphics[width=1.75\columnwidth]{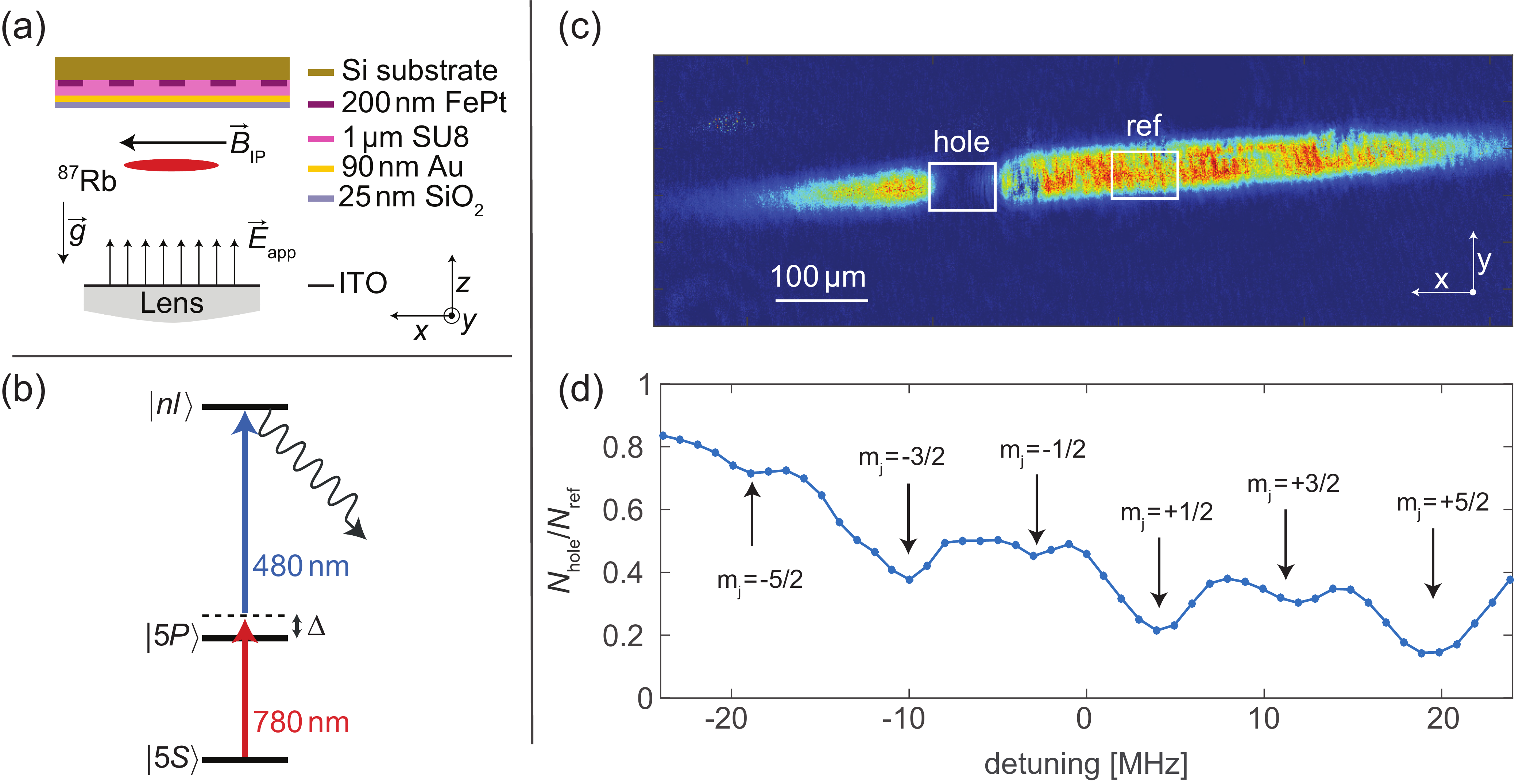}
\caption[]{
(a) Schematics of the apparatus (not to scale). 
The Si atom chip and the fabricated layers are shown \cite{Leung:2014gw}. The atom chip is facing downwards, and
the atomic cloud is magnetically trapped $\sim100\,\mu$m below the surface, in a Ioffe-Pritchard field $B_\mathrm{IP}$ in the $x$-direction.
The in-vacuum lens is used for high-resolution imaging (19$\,$mm from the surface, NA=0.4, and resolution of $\sim1\,\mu$m), and it is covered with Indium-Tin Oxide (ITO) such that a homogeneous electric field in the $z$-direction, $E_\mathrm{app}$, can be applied between the lens and the Au layer of the atom chip.
(b) Level diagram of the two-photon excitation scheme with 780$\,$nm and 480$\,$nm laser light. The detuning $\Delta/2\pi$ is chosen to be $1.5\,$GHz.
(c) An example of an optical density image of the cloud at $\sim170\,\mu$m, where the position of the excitation laser is visible as a hole in the cloud.  
The fine substructure in the cloud is due to imperfections in the atom chip surface and fringes in the imaging laser.
(d) A sample spectrum of the 23$D_{5/2}$ state at a distance of $\sim170\,\mu$m in a near electric-field-free environment achieved after exposing the chip surface to UV light, see Sec. \ref{sec:UV} for details.
The ratio of the number of atoms in the ``hole" area ($N_\mathrm{hole}$) to a ``reference" area ($N_\mathrm{ref}$), both marked in the cloud image in (c), is plotted as we scan the frequency of the 480$\,$nm laser.
Note that this ratio does not reach 1 because $N_\mathrm{hole}<N_\mathrm{ref}$, even without excitation lasers.
The detuning is relative to the transition frequency as expected from electromagnetically induced transparency measurements in a room temperature vapour cell without electric and magnetic fields.
This zero-detuning frequency is kept throughout this paper.
The spacing between the different transitions is $7.4\pm0.3\,$MHz, corresponding to $B_\mathrm{IP}=4.4\pm0.2\,$G at the trap bottom. From independent measurements of the trap bottom we get $B_\mathrm{IP}=4.2\pm0.1\,$G.}
\label{fig:ODimage}
\end{figure*}

The experiments are performed in an atom chip setup that features a micro-structured FePt layer for the creation of magnetic micro-trap potentials \cite{Gerritsma:2007vo,Whitlock:2009eg}. This layer is covered with an Au and a SiO$_2$ coating, which is acting as a mirror for a mirror magneto-optical trap (MOT) configuration \cite{Reichel:1999ho}. 
Beneath the Au there is an additional 1$\,\mu$m thick layer of SU8 polymer.
A schematic of the apparatus is shown in figure \ref{fig:ODimage}(a).
The atoms are collected in the MOT from a background vapor of Rb atoms. They are then optically pumped into the $|F,m_F\rangle=|2,2\rangle$ Zeeman sublevel, and transferred into a magnetic Ioffe-Pritchard (IP) trap \cite{Pritchard:1983dd} formed by a z-shaped wire located behind the Si substrate. 
We cool the cloud by forced RF evaporation to $\sim30\,\mu$K and position it at different distances of 10-200$\,\mu$m below the atom chip by varying the bias magnetic field. 
The cloud is then exposed for $100\,\mu$s to two laser beams with wavelengths 480$\,$nm and 780$\,$nm both having $\sim100\,\mu$m beam waist ($1/e^2$ radius), which is smaller than the axial size of the cloud but comparable to the radial size. We use a two-photon process with a detuning of $\Delta/2\pi=+1.5\,$GHz from the intermediate 5$P_{3/2}$ level, see figure \ref{fig:ODimage}(b) for a level diagram. 
The laser powers used are 140$\,$mW and 80$\,\mu$W respectively.
The beams propagate in the $z$-direction (perpendicular to the atom chip surface), and with linear polarisation, so that both contain $\sigma^+$ and $\sigma^-$ polarisation with respect to the magnetic field at the trap minimum $B_{\mathrm{IP}}$, which is tilted by a few degrees from the $x$-direction.
Both beams are reflected back by the Au surface, such that they overlap with the incoming beam.
The two-photon transition excites the exposed atoms to either the 23$D_{5/2}$ or 25$S_{1/2}$ state, from which they decay to non-detectable states or are ionised, such that they appear as lost from the absorption image. The image is taken shortly ($100\,\mu$s) after the excitation pulse to prevent neighbouring atoms to refill the depleted area.
See figure \ref{fig:ODimage}(c) for a sample optical density image of the cloud immediately after the excitation pulse showing the lost atoms as a hole, and the tilt between $B_\mathrm{IP}$ and the $x$-direction. 
The typical duration of one experimental cycle is $\sim20\,$s. The number of atoms in the depleted area is normalised against a non-influenced reference area in the cloud. This greatly suppresses noise from shot-to-shot fluctuations of the overall number of atoms and improves the visibility of the spectrum. Both lasers are frequency stabilised to a high-finesse cavity using a sideband-locking scheme with typical linewidths of less than 10$\,$kHz \cite{Naber:2015wk}. We scan either the frequency of the 480$\,$nm laser by changing the sideband frequency of the locking mechanism, or the frequency of the 780$\,$nm light by an acousto-optical modulator (AOM). The first method is used in figure~\ref{fig:2photon_muTraps}, whereas the latter is used for all the other measurements in this paper. We measure the spectrum of the excited sublevels by taking absorption images at different excitation frequencies. Figure~\ref{fig:ODimage}(d) shows an example of such a spectrum in a near electric-field-free environment ($<1\,$V/cm). Our light polarisation allows for transitions with $\Delta m_J=0,\pm2$, so that starting from the stretched ground state ($m_J=1/2$) we address 23$D_{5/2}$ with $m_J=-3/2,1/2,5/2$. The transitions at $m_J=-5/2,-1/2, 3/2$ are also visible, due to imperfect polarisation and remaining electric fields, albeit suppressed relatively to the others.

\section{Results}

\subsection{Measuring stray electric fields}

\begin{figure}[t]
\centering
\includegraphics[width=\columnwidth]{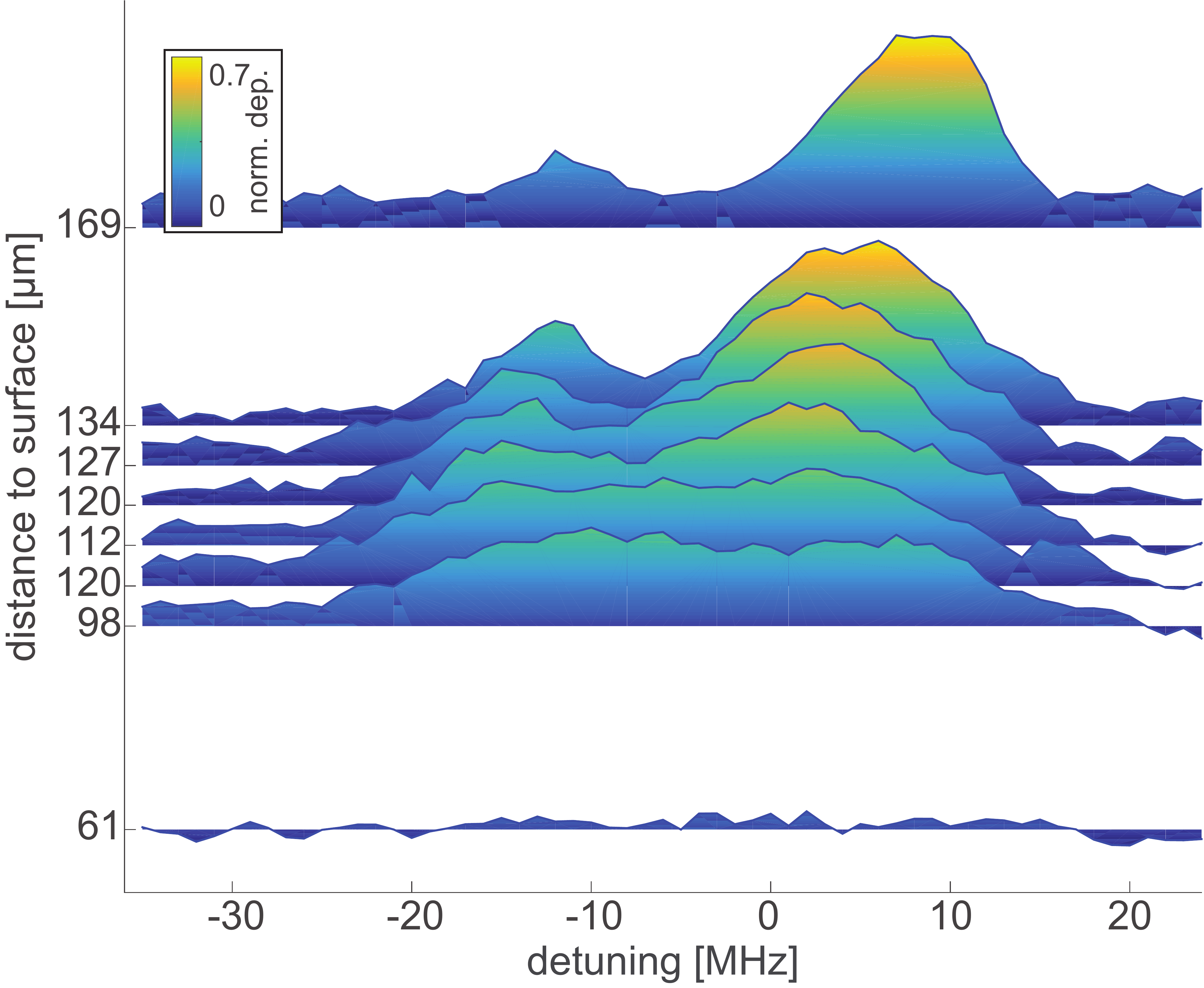}
\caption[]{
Depletion spectra for the $23D_{5/2}$ Rydberg state at different distances from the atom chip surface.
We plot the normalised depletion  $1-\frac{N_\mathrm{hole}}{R\,N_\mathrm{ref}}$ as the colour scheme (colour bar is linear here and throughout the paper), where $R=0.77$ is the ratio between ${N_\mathrm{hole}}$ to $N_\mathrm{ref}$ without the excitation lasers.
The distance to the surface, represented as vertical shift between different curves, is taken from magnetic potential calculations which are calibrated to the experiment.
Already at a distance of 169$\,\mu$m there is a shift of $\sim-10\,$MHz compared to the transition frequency presented in figure \ref{fig:ODimage}(d).
In addition, comparing  the absorption at $134\,\mu$m to $98\,\mu$m the spectrum almost completely disappears after a change of only $36\,\mu$m, together with a significant broadening.
}
\label{fig:heightScan}
\end{figure}

The future goal of the experiment is to excite atoms to a Rydberg state in traps only $10\,\mu$m from the atom chip surface \cite{Leung:2014gw,Leung:2011el}, which requires a good knowledge of the local stray electric fields. We probe these fields by positioning the cloud at different distances to the surface and measuring the excitation spectrum to the 23$D_{5/2}$ state. Figure \ref{fig:heightScan} shows these spectra at different distances. 
Upon approaching the surface, the depletion peaks are shifted to negative frequencies, and a reduction and broadening of the spectrum is visible. Both these effects suggest increasing electric fields and increasing electric field gradients with decreasing distance to the surface.

\begin{figure}[t]
\centering
\includegraphics[width=\columnwidth]{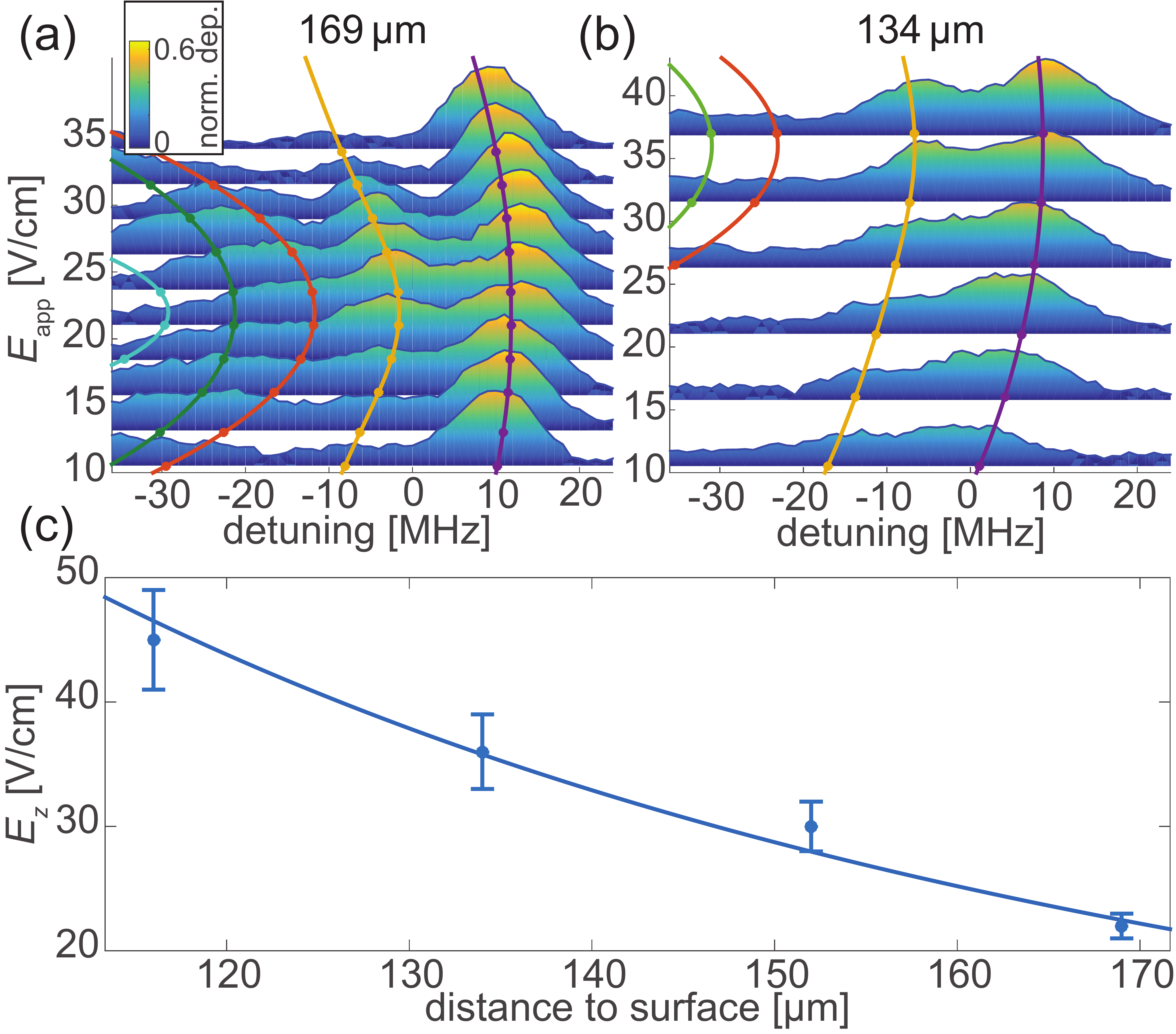}
\caption[]{Depletion spectrum of the $23D_{5/2}$ Rydberg state taken at distances of (a) 169$\,\mu$m and (b) $134\,\mu$m  from the surface for different applied fields ${E_\mathrm{app}}$ in the $z$-direction, showing a clear dependency of the overall frequency shift of the spectrum on the applied field. The solid lines are theoretical predictions for the peak positions based on the method described in the text, where the stray field parameters have been chosen to best match the observed spectral dependency on ${E_\mathrm{app}}$. (c) Extracted stray field values in the $z$-direction, together with a fit based on the electric field of a Gaussian patch of Rb ad-atoms.}
\label{fig:noLeds}
\end{figure}

To better examine the underlying mechanism, we use the fact that we can compensate the electric field ${E_\mathrm{z}}$ in the $z$-direction by applying a homogeneous field ${E_\mathrm{app}}$ using the in-vacuum lens covered with Indium-Tin Oxide (ITO). The minimum frequency shift of the spectrum occurs when ${E_\mathrm{z}}$ is effectively cancelled by ${E_\mathrm{app}}$, so we take spectra for different applied electric fields.
Figure~\ref{fig:noLeds}(a,b) shows two measurements for 23$D_{5/2}$ at two different heights.
To extract realistic field values from these measurements, we evaluate the atomic transition frequencies for a given magnetic $B_\mathrm{IP}$ and electric field. The magnetic field is measured independently and is directed predominantly in the $x$-direction [we neglect the small tilt between $B_\mathrm{IP}$ and the $x$-direction visible in figure~\ref{fig:ODimage}(c)]. We assume a linear Zeeman shift for the $m_J$ states and write the Zeeman Hamiltonian for a quantisation axis defined by the total electric field, constituted by (${E_\mathrm{x}},{E_\mathrm{y}},{E_\mathrm{z}}-{E_\mathrm{app}}$). The Stark shifts are taken from the diagonalisation of the Stark-Hamiltonian in a sufficiently large set of Rydberg states that are close in energy. 
Subsequently we diagonalize the combined Zeeman-Stark Hamiltonian. 
If we plot the atomic energies as a function of the applied field ${E_\mathrm{app}}$, we can well reproduce the peak position in the experimental spectra for a given set of stray electric fields (${E_\mathrm{x}},{E_\mathrm{y}},{E_\mathrm{z}}$). For example, the measurement in figure \ref{fig:noLeds}(a) yields (${E_\mathrm{x}},{E_\mathrm{y}},{E_\mathrm{z}}$)=(5,8,22)$\,$V/cm. We find a similar value for $E_\mathrm{z}$ by an independent measurement for 25$S_{1/2}$. 

The extracted stray field values $E_\mathrm{z}$ are plotted in figure \ref{fig:noLeds}(c), which shows a strong increase in electric field from $22\,$V/cm to $40\,$V/cm with decreasing distance to the surface. The observed fields are about one order of magnitude larger than what was found in a previous chip experiment with a plain Au-surface \cite{Tauschinsky:2010ep}. Our measurement shows that the stray fields' $z$-component points away from the surface. 
This is consistent with a field induced by Rb ad-atoms \cite{Tauschinsky:2010ep,Sedlacek:2015tg}.
The electric field above the centre of a Gaussian patch of Rb ad-atoms can be described in the $z$-direction as  \cite{Tauschinsky:2010ep}
\begin{equation}
E_\mathrm{z}=\frac{d_0}{2w\epsilon_0}\times\left[-Z+e^{\frac{1}{2}Z^2}\sqrt{\frac{\pi}{2}}(1+Z^2)\erfc\left(\frac{Z}{\sqrt{2}}\right)\right]
\label{eq:patch}
\end{equation}
with the $e^{-1/2}$ patch radius $w$, $d_0$ the peak dipole density, $\erfc$ the complementary error function, and $Z=z/w$. When we fit this expression to our data in figure \ref{fig:noLeds}(c), we retrieve a patch radius $w=70\,\mu$m and a peak dipole density $d_0=1.2\times10^7\,$Debye/$\mu$m$^2$. Assuming a dipole moment of 12 Debye per ad-atom \cite{Sedlacek:2015tg}, we retrieve an average ad-atom spacing of $1\,$nm, which is comparable to the value found in  \cite{Sedlacek:2015tg}, but an order of magnitude smaller than what was found in \cite{Tauschinsky:2010ep}. The gradient at a height of $134\,\mu$m above the centre of the patch is $\partial E_\mathrm{z} /\partial z =6200\,$V/cm${}^2$.

\begin{figure}[t]
\centering
\includegraphics[width=\columnwidth]{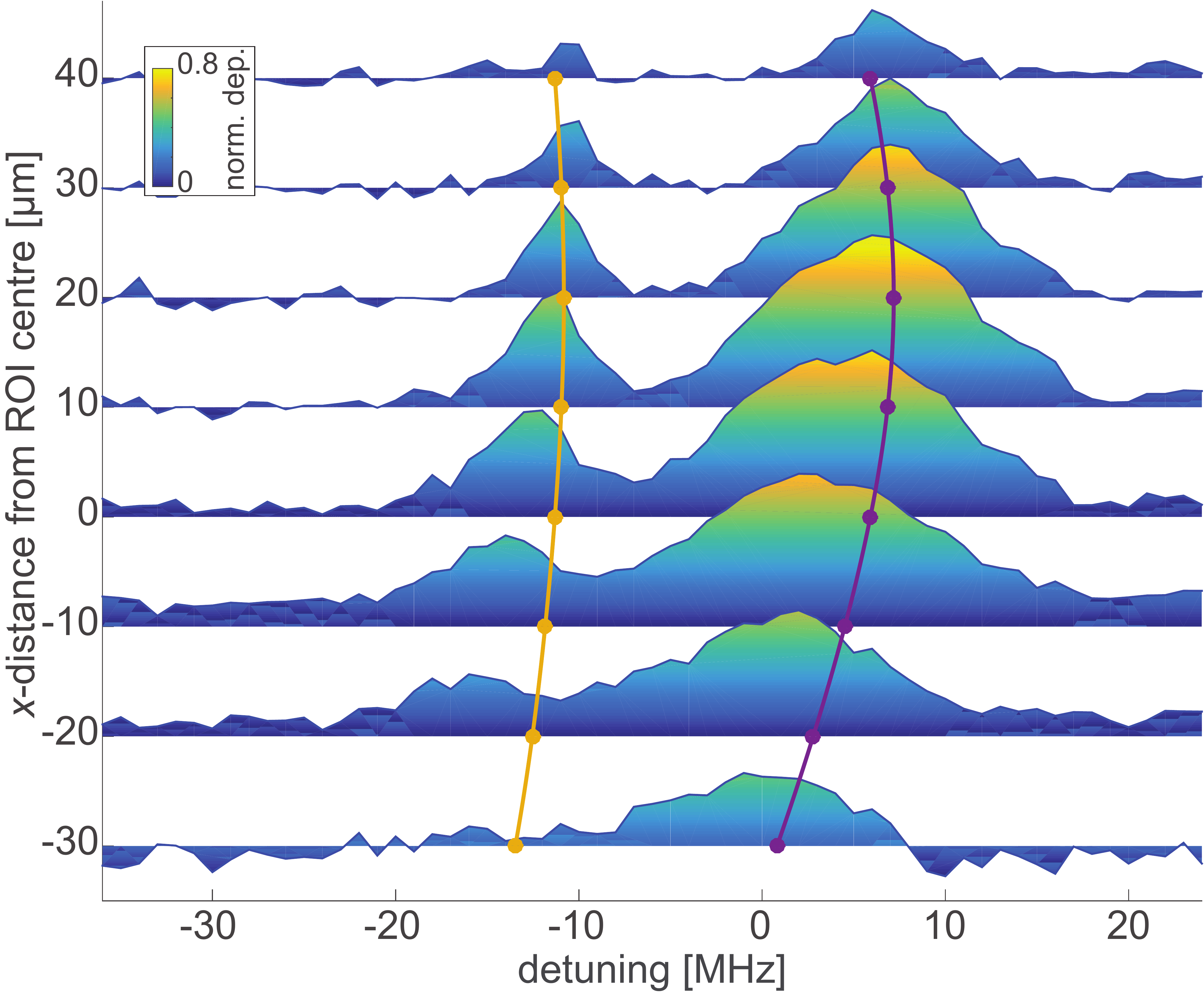}
\caption[]{
Depletion spectra of the $23D_{5/2}$ Rydberg state for equally spaced, $10\,\mu$m wide sub-areas along the $x$-direction in the area marked as ``hole'' in figure \ref{fig:ODimage}(c).
The strong position dependence indicates a strong electric field gradient in the $x$-direction. The theoretical lines assume a linear field gradient in the $x$-direction. 
}
\label{fig:ROIregions}
\end{figure}

In order to estimate the $x$-component of the electric field gradient, we further analyse the measurement in figure \ref{fig:heightScan} taken at a distance of 134$\,\mu$m. In this measurement the $480\,$nm laser was pulsed for $200\,$ms (see Sec. \ref{ch:temp}) and the $780\,$nm laser for $100\,\mu$s. Instead of defining a region-of-interest (ROI) that includes the entire excitation beam area [the ``hole'' square in figure \ref{fig:ODimage}(c)], we divide this ROI into smaller areas in the $x$-direction, each $10\,\mu$m wide, and plot the spectrum for the different sub-areas (figure \ref{fig:ROIregions}). A strong change in transition frequency is visible over only 100$\,\mu$m. In order to extract the corresponding field gradient, we use the theoretical description of the atomic energy levels assuming a linear field gradient in the $x$-direction.
From that we can extract a gradient value of $\partial E/\partial x=3500\,$V/cm${}^2$, comparable to the value found for the $z$ field gradient. We do not observe a gradient in the $y$-direction, possibly because we are naturally limited by the smaller cloud size in that direction.

\begin{figure}[t]
\centering
\includegraphics[width=\columnwidth]{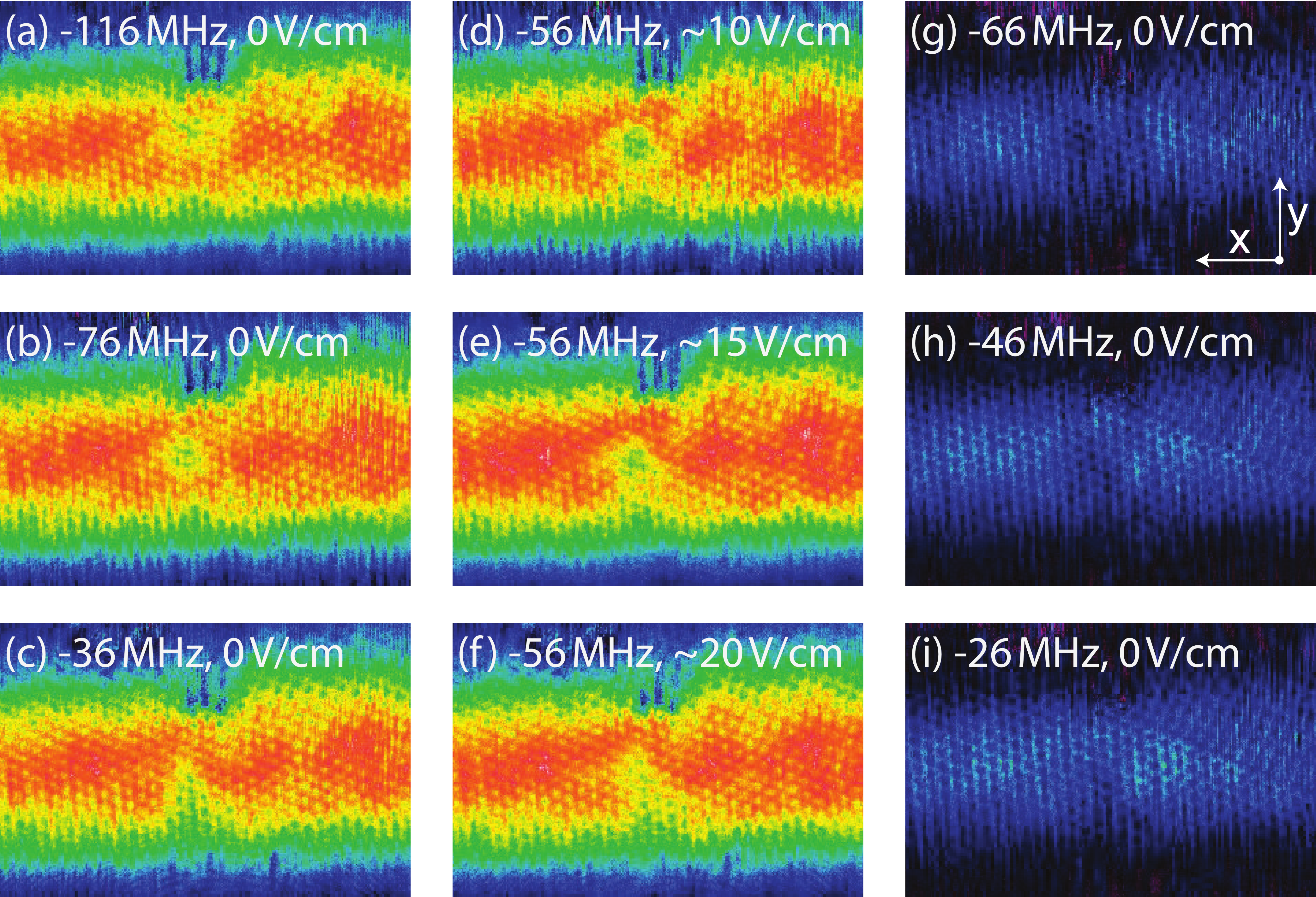}
\caption[]{
Optical density images normalized to the same optical density value close to the atom chip surface, such that the magnetic micro-traps are influencing the magnetic trapping potential. Atoms which are excited to the $25S_{1/2}$ Rydberg state are lost and appear as a hole in the cloud. Here, the detuning is relative to an independent electromagnetically induced transparency measurement. 
(a-c) Changing the frequency of the 480$\,$nm excitation laser changes the position that matches the excitation frequency, due to electric field gradients.
(d-f) Changing $E_{\mathrm{app}}$ changes the excitation position.
Images (a-f) were taken with the cloud $\sim15-20\,\mu$m from the surface.
The main potential is supplied by a macroscopic z-shaped wire, but the lattice created by the permanent magnet layer on the atom chip is starting to be visible.
(g-i) Exciting to Rydberg state while the atoms are trapped by the magnetic lattice, only $\sim10\,\mu$m from the surface. 
The field-of-view of all the images is 300 by 200$\,\mu$m in the $x$- and $y$-directions, respectively.
}
\label{fig:2photon_muTraps}
\end{figure}

Our atom chip features a micro-structured FePt layer beneath the Au coating, which is used together with an external bias field to create an array of magnetic micro-traps close ($\sim10\,\mu$m) to the surface \cite{Leung:2014gw}.
In order to obtain data close to the surface ($\sim10\,\mu$m), where the magnetic trapping potential is influenced by the magnetic micro-traps, we change our excitation scheme to obtain a higher Rabi frequency.
Contrary to the rest of the measurements in this paper, we tune our lasers on resonance ($\Delta=0\,$MHz) with the intermediate 5$P_{3/2}$, $F=3$ level, leading to a two-photon on-resonance  transition with much higher Rabi frequency to the 25$S_{1/2}$ state. The imaging laser is used here for the excitation to the intermediate level at $780\,$nm with a pulse time of $100\,\mu$s. The $480\,$nm laser is on for $200\,$ms. Figure \ref{fig:2photon_muTraps} shows optical density images while changing the frequency of the 480$\,$nm laser or the applied electric field $E_{\mathrm{app}}$.
As in figure~\ref{fig:ODimage}(c) atoms excited to a Rydberg state are lost and appear as a hole in the cloud. The size of this hole is much smaller than the laser beam diameters. The hole thus marks where the excitation is resonant. We use two different trap configurations: a macro-wire trap where the lower part of the atomic cloud is already influenced by the micro-trap potential [figure \ref{fig:2photon_muTraps}(a)-(f)], and atoms confined in the micro-traps [figure \ref{fig:2photon_muTraps}(g)-(i)]. In those images, a change in excitation frequency results in a spatial translation of the depletion area in the $y$-direction.
Similarly, a change in applied field $E_{\mathrm{app}}$ leads to a spatial shift in the same direction. This points at strong electric field gradients in the $y$-direction. Remarkably, the transition frequency changes by $100\,$MHz over a distance of a few tens of micrometers, corresponding to a gradient of about $\partial |E|/\partial y=1700\,$V/cm${}^2$. Varying the applied voltage on the lens, the depletion area traverses only half of the cloud, which suggests that we only cancel the $E_{\mathrm{z}}$ with strong fields $E_{\mathrm{x}},E_{\mathrm{y}}$ still present. For our measurements at that small distance, the spatial depletion changes from shot to shot, and the involved frequencies from day to day, making a quantitative analysis unreliable.
\begin{figure}[t]
\centering
\includegraphics[width=\columnwidth]{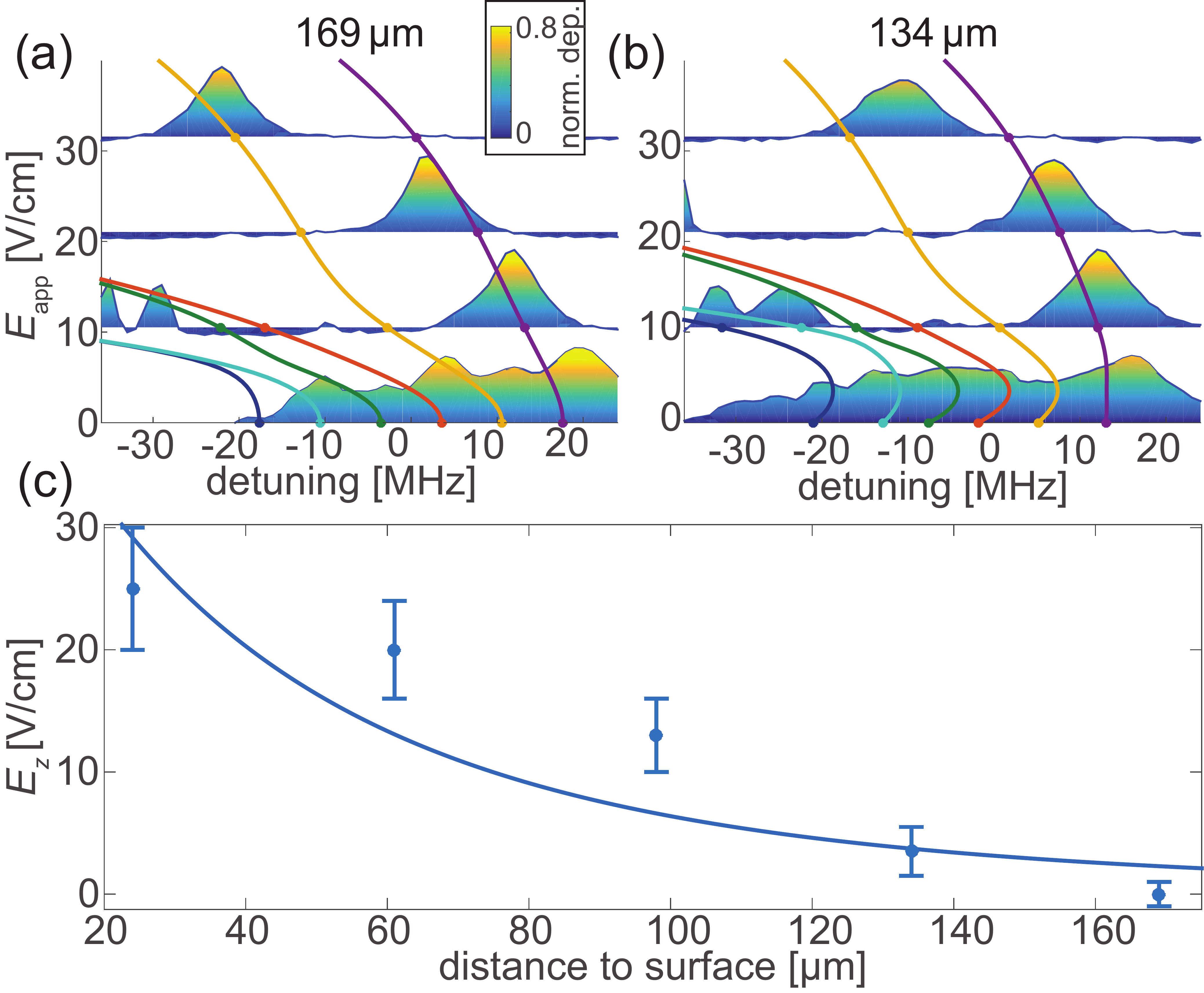}
\caption[]{
Normalised depletion spectra of the $23D_{5/2}$ Rydberg state at two heights [(a): 169$\,\mu$m and (b): 134$\,\mu$m] and with different applied external electric fields while using the LEDs (LEDs have been switched on for the first time on the day  this data was taken).
The spectrum for $E_\mathrm{app}=0$ in (a) is shown in figure \ref{fig:ODimage}(d) as well, and indicates small ($<1\,$V/cm) electric stray field at that height. A clear reduction in the stray electric field is visible after using the UV light, but the field does not vanish completely. (c) The extracted electric stray field value $E_\mathrm{z}$ in $z$-direction plotted against the distance to the surface. Compared to figure \ref{fig:noLeds}(c) the field is strongly reduced.
}
\label{fig:LEDsEffect}
\end{figure}

\subsection{Reduction of the stray electric fields}\label{sec:UV}
The presence of strong stray electric fields and gradients near the surface poses an obstacle for exciting Rydberg atoms at the chip. Reducing these fields is crucial to our goal, and we investigate several methods for removing the source of these fields. One method, as suggested in \cite{Sedlacek:2015tg}, is to excite all the atoms in the MOT to a Rydberg state, which will lead to subsequent ionisation of a large fraction of the excited atoms. The resulting charges can settle on the surface and compensate for the electric stray field. Contrary to the findings in \cite{Sedlacek:2015tg}, where a low number of electrons significantly reduces the positive ad-atom field, we do not see any compensation effect. Looking at the differences between the experiments, a possible explanation might be that we do not have a bulk mono-crystalline layer of SiO$_2$, and that the exposure of the surface to the excitation lasers might remove the free charges.

In a second attempted method we deliberately deposit large atomic clouds on the surface, as it was found in \cite{HermannAvigliano:2014bc} that a more homogeneous layer of adsorbed atoms can decrease the stray fields. In our case, however, this procedure increases the stray fields. 

As a third method, we illuminate the atom chip with UV light at $365\,$nm as we expect UV light to influence the surface ad-atoms via the light-induced atomic desorption (LIAD) effect \cite{Meucci:1994gx,TorralboCampo:2015bo}. The UV light is provided by an array of nine $1\,$W LEDs, which is brought in close proximity to the vacuum quartz cell with the UV light partially collimated towards the chip surface. The LEDs are switched on during the entire MOT stage of the experimental procedure, but switched off before the magnetic trapping to reduce the background pressure and increase the in-trap lifetime. The UV light results in a large increase in the number of atoms, due to LIAD from the vacuum quartz cell walls, releasing a lot of additional Rb atoms. To quantify the effect on the electric stray fields, we perform the same measurements as in figure~\ref{fig:noLeds}. The results are shown in figure~\ref{fig:LEDsEffect}(a,b), and show a spectrum with small electric stray field at 169$\,\mu$m, as indicated by the equally spaced depletion peaks. As shown in figure~\ref{fig:ODimage}(d), the level spacing is consistent with the Zeeman splittings. If we decrease the distance, we find an increasing electric field. In general, even though the data is less well reproduced by our theory after using the LEDs, the $E_{\mathrm{z}}$ value can still be estimated well (compared to $E_{\mathrm{x}}$,$E_{\mathrm{y}}$) from the spectrum. The field in the $z$-direction and a fit to a Gaussian patch model (Eq.~\ref{eq:patch}) are plotted in figure \ref{fig:LEDsEffect}(c), showing a strong reduction compared to figure \ref{fig:noLeds}(c), although the fields are still larger than those of reference \cite{Tauschinsky:2010ep}. We can also conclude from the poor fit that the Gaussian patch model no longer represents the underlying ad-atom distribution.

\subsection{Temporal behaviour of the stray fields}
\label{ch:temp}
\begin{figure}[t]
\centering
\includegraphics[width=\columnwidth]{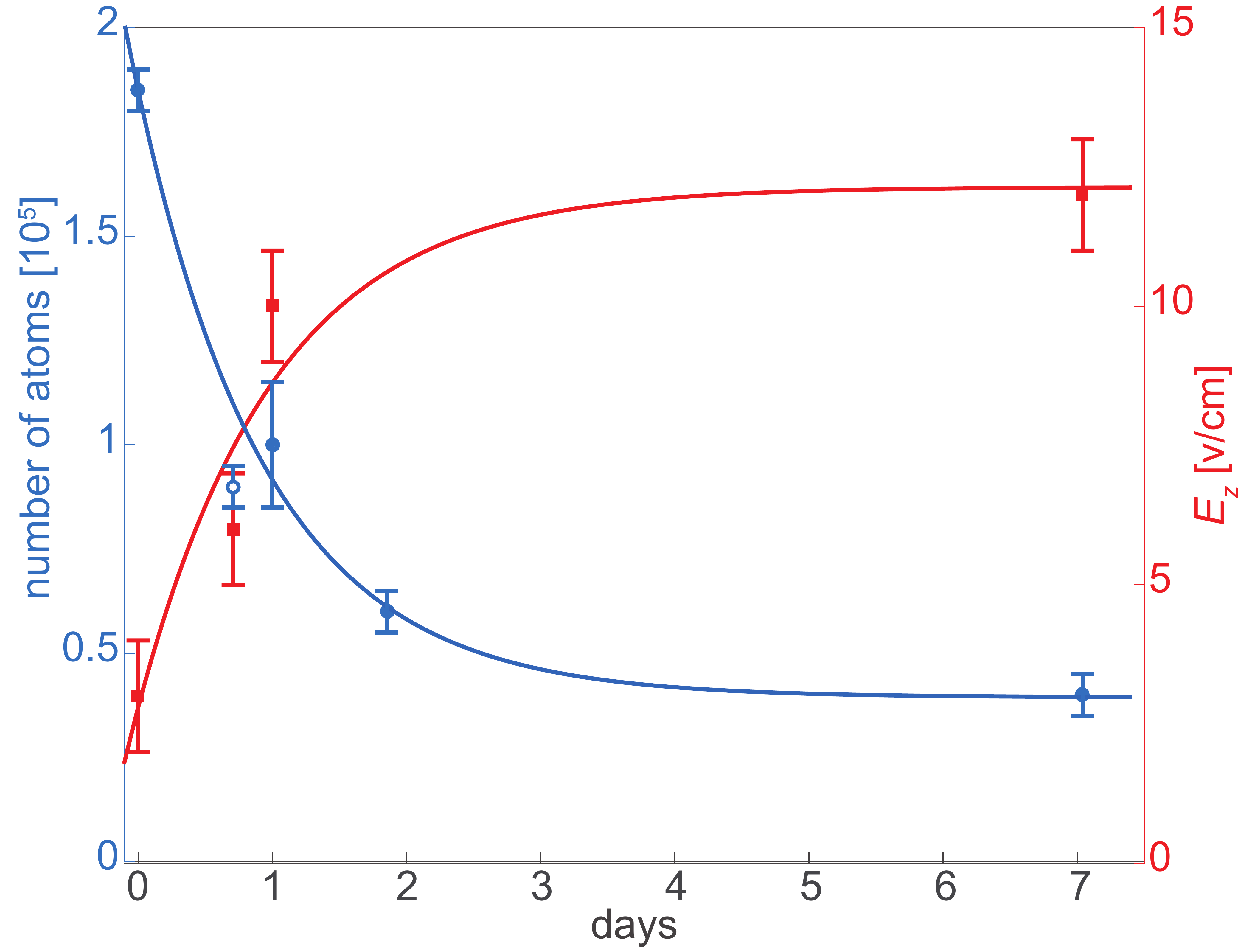}
\caption[]{
Number of atoms and electric stray fields in the first week of using the UV LEDs at a distance of 134$\,\mu$m to the surface.
The electric stray field is measured by applying an external field, \E, and measuring the excitation spectrum,  as in figure \ref{fig:LEDsEffect}. 
The decay time of the number of atoms is equal to the increase time of the stray fields, $\sim$1 day.
The data marked with an open blue circle is not included in the number of atoms fit because it belongs to a dataset with a colder, hence smaller, cloud.
The corresponding measurement of the stray fields is included in the stray fields fit because the field measurement is independent of the number of atoms.
}
\label{fig:LEDsWithTime}
\end{figure}
In order to better understand the influence of the UV light on the stray fields, we look at the temporal dynamics of this effect. Within one day of illuminating the cell and the atom chip with UV light continuously, leading to a strong initial reduction of stray fields, the number of atoms starts to decrease and simultaneously the stray fields increase. 
The reduction of the number of atoms indicates that the UV illumination cleans the inside of the vacuum quartz cell from Rb atoms accumulated there. However, it is unclear why this leads to an increase of the stray fields.
Figure \ref{fig:LEDsWithTime} shows the reduction of the number of atoms after the evaporation stage during the first week of operating the LEDs, and the corresponding increase of the stray electric field at a height of 134$\,\mu$m. Both exponentials have the same time constant of $\sim1$ day, suggesting a negative correlation between number of atoms and stray fields under our operating conditions. After a week of operating the LEDs, the decrease in number of atoms does not leave sufficient atoms in the magnetic trap for taking spectra. When we increase the background pressure again by switching off the LEDs and operating the system normally for a couple of hours, we can retrieve the original number of atoms. If we then switch on the LEDs we see again the same reduction of stray fields, with a subsequent drop in number of atoms and increase in stray fields.
\begin{figure}[t]
\centering
\includegraphics[width=\columnwidth]{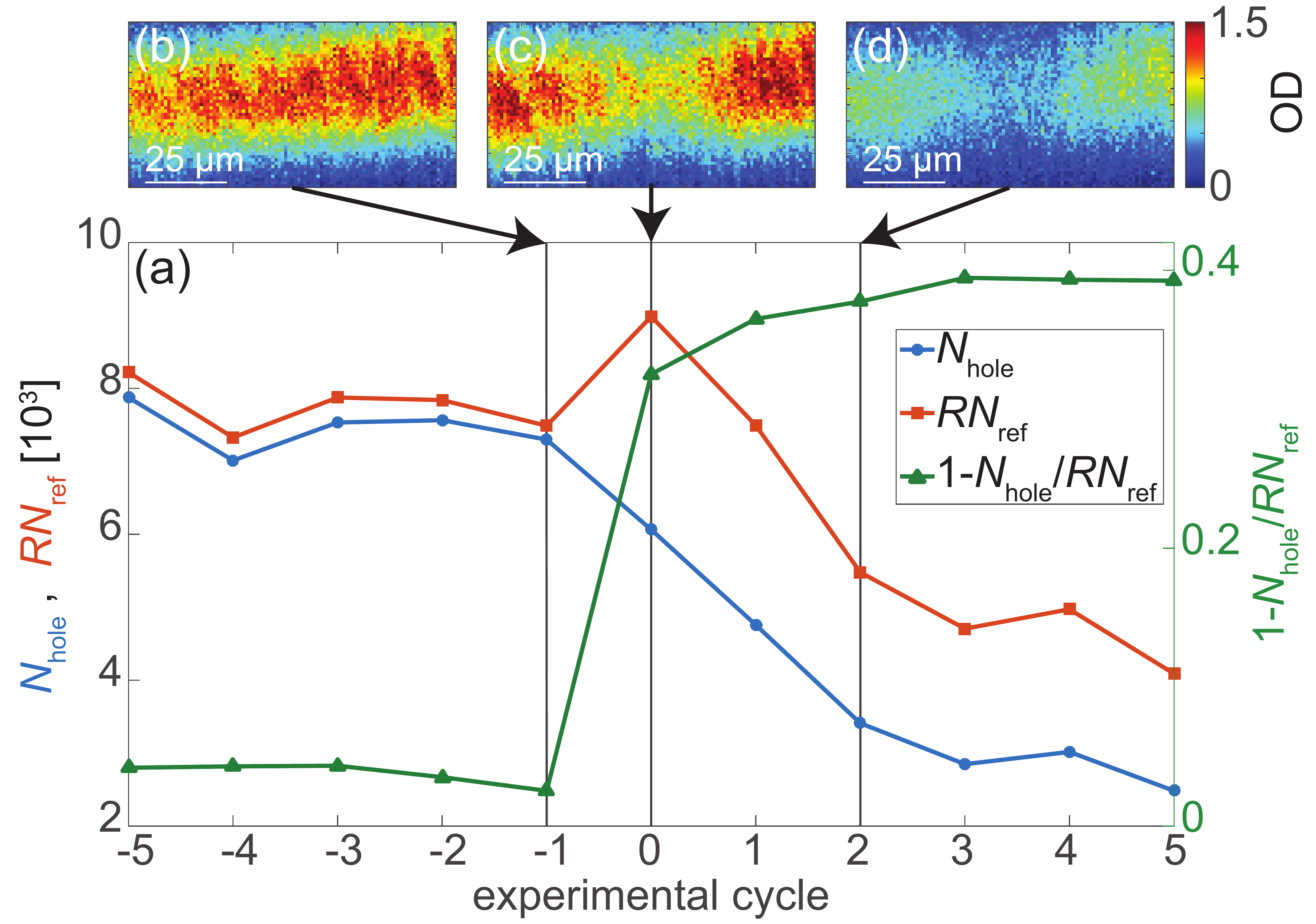}
\caption[]{
Temporal behaviour of the stray field as the LEDs are turned on.
(a) The excitation lasers are on and tuned to the field-free frequency [detuning of 20$\,$MHz, as shown in figure \ref{fig:ODimage}(d)].
Experimental cycle 0 defines the first cycle with LEDs turned on.
In that cycle we observe a small increase in the number of atoms with corresponding increase in the ratio $1-N_\mathrm{hole}/RN_\mathrm{ref}$, indicating the fast change of the stray fields.
(b-d) Optical density images (field-of-view of 100 by 50$\,\mu$m) corresponding to three experimental cycles: (b) the image taken before turning on the UV LEDs, (c) the first image taken after turning on the UV LEDs, and (d) an image taken after a few experimental cycles showing the reduction in the total number of atoms \emph{without} a corresponding increase in the stray fields.
}
\label{fig:LEDsOn}
\end{figure}
In order to show that the reduction of stray electric fields due to UV light occurs within one experimental cycle ($\sim20$ s) after switching on the LEDs, we use the following procedure: (i)  build up  Rb pressure by normally operating the system without LEDs, usually for several hours, (ii)  tune the excitation  lasers to the low-field transition frequency as measured in figure \ref{fig:LEDsEffect}, yielding no depletion at all, and (iii) turn on the UV light. In this procedure, the depletion becomes visible only if the stray fields decrease. Figure \ref{fig:LEDsOn} is the result of such a procedure, where the LEDs are turned on in the beginning of experimental cycle 0. An immediate reduction of the stray fields in one experimental cycle is visible.
After a few more experimental cycles we had to stop the measurement in order to rebuild the Rb pressure.

\begin{figure}[t]
\centering
\includegraphics[width=\columnwidth]{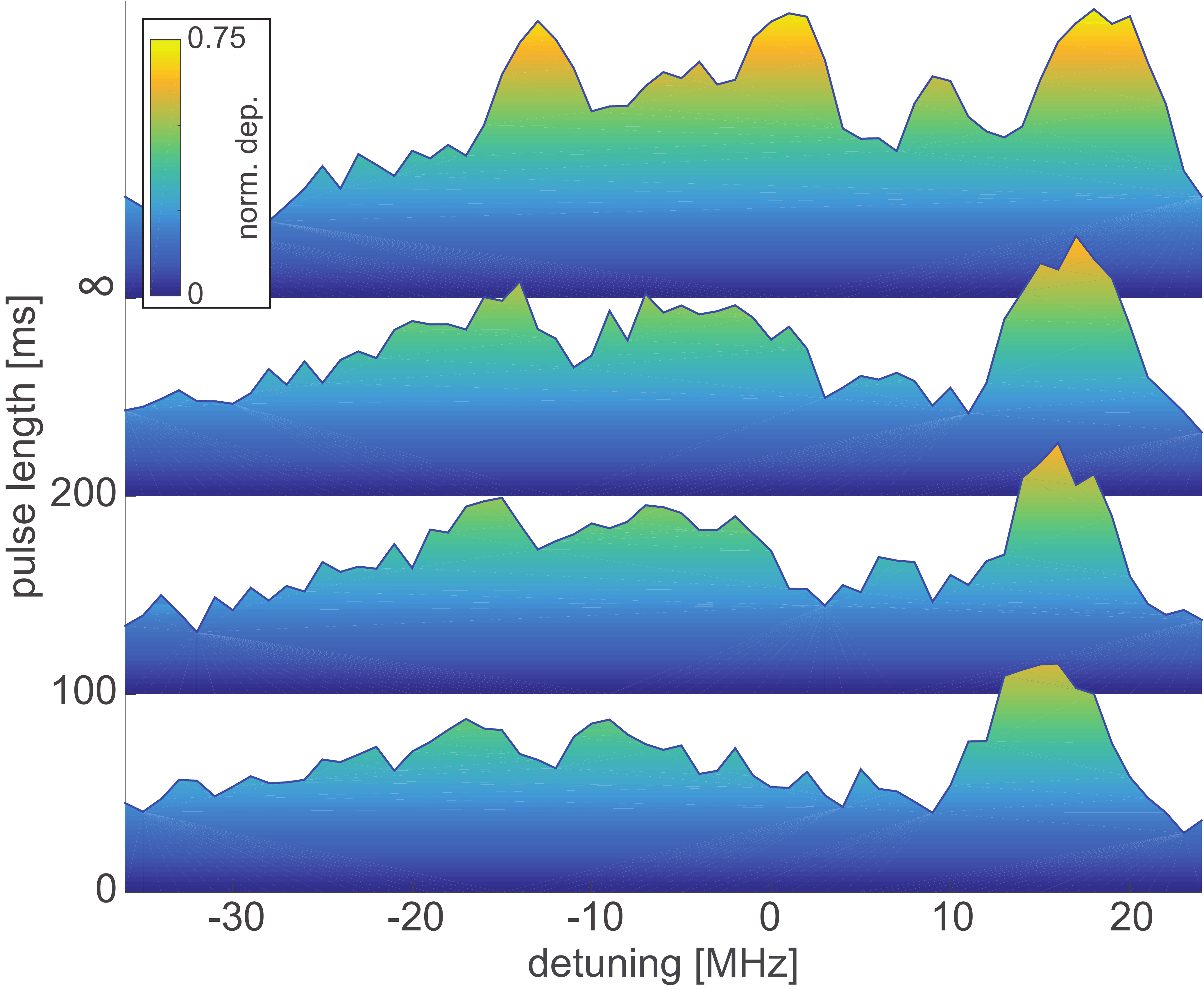}
\caption[]{
Depletion spectra of the $23D_{5/2}$ Rydberg state for varying pulse times of the 480$\,$nm laser. The laser illuminates the atom chip surface before the 780$\,$nm excitation laser creates Rydberg atoms. The four spectra are taken for (from bottom to top): no additional 480$\,$nm pulse, 100$\,$ms and 200$\,$ms long pulses, and continuously  open (marked with $\infty$). With increasing pulse length the spectrum is shifted to more positive frequencies and more peaks can be discriminated, indicating smaller stray fields.
}
\label{fig:blueLaserTime}
\end{figure}

The temporal dynamics shown in figure~\ref{fig:LEDsWithTime} suggest that the 
dominant effect of the UV light is not the removal of ad-atoms from the surface itself, 
as the stray fields increase again while the UV light is present. 
We speculate that we influence the ad-atom distribution by increasing the Rb background pressure thereby creating a more uniform or larger layer, resulting in a decrease of stray fields. 

We also examine the influence of the 480$\,$nm excitation laser on the stray fields. The laser hits the atom chip surface at normal incidence with a $\sim40\,\mu$m 1/e$^2$ radius. We estimate that this increases the local surface temperature by $\sim50\,$K  (this number is based on our laser intensity, surface reflectivity, and the thermal conductivities of the Au film, the SU8 layer, and the Si substrate).
This temperature increase can cause desorption of ad-atoms which reduces stray fields as shown in reference \cite{Sedlacek:2015tg}. In addition, we expect an effect due to LIAD, as the local light intensity is orders of magnitude higher than for the LED array. If the surface is exposed for several seconds to the laser \emph{after} the excitation pulse and the imaging sequence, there is no change of the stray fields in the next shot ($\sim15$s delay). On the other hand, if the surface is illuminated \emph{before} the excitation pulse, we measure a reduction of the electric stray field, which increases with the exposure time, as shown in figure \ref{fig:blueLaserTime}.

This reduction is also present in figure~\ref{fig:ROIregions} and \ref{fig:2photon_muTraps} compared to figure~\ref{fig:noLeds}, as those measurements were taken with 200$\,$ms long 480nm pulses. The minimum stray field in the $x$-direction in figure~\ref{fig:ROIregions} roughly corresponds to the beam centre. Furthermore, in figure~\ref{fig:2photon_muTraps}, the depletion is only visible in the beam centre. The electric stray fields corresponding to the detuning of $(-116,-76-36)\,$MHz in figure~\ref{fig:2photon_muTraps}(a-c) are $(25,20.25,14)\,$V/cm, which is significantly lower than the fields in figure \ref{fig:noLeds}(c).

\section{Discussion}

In reference \cite{McGuirk:2004dm} electric surface fields were measured $\sim10\,\mu$m from different surfaces by a change in oscillation frequency of a Bose Einstein condensate. The fields were attributed to Rb ad-atoms on the surface. When BK7 glass is used as surface material, the stray fields are significantly lower compared to Si or Ti surfaces, less than $1\,$V/cm. Similarly low electric fields were found in \cite{Harber:2005jl} where fused silica was used as surface material. A low field was also found in micro-metre size vapor glass cells \cite{Kubler2010Coherent}.
In contrast to this, measurements with Yttrium \cite{Obrecht:2007br} and Au \cite{Tauschinsky:2010ep,Hattermann:2012ho} show larger stray fields which increase with the amount of deposited Rb on the surface. In comparison, this seems to suggest that a dielectric surface is superior to a metal surface in terms of stray fields. However, for our SiO$_2$ coating we find stray fields which are roughly one order of magnitude larger than for our earlier atom chip with an Au surface \cite{Tauschinsky:2010ep}. Our finding is consistent with the measured stray fields found in \cite{Sedlacek:2015tg} for SiO$_2$. Furthermore, as the sign of the electric dipole moment can be used to distinguish between chemisorbed and physisorbed ad-atoms \cite{Chan:2014kb}, our measurements supports chemisorbed ad-atoms compared to a weaker binding due to van der Waals forces as assumed in \cite{McGuirk:2004dm}. This again is consistent with \cite{Sedlacek:2015tg}.

In contrast to the measurements in \cite{Sedlacek:2015tg}, we do not see a decrease of stray fields when producing free charges by Rydberg excitation in the MOT. This might be related to two differences in the experiments: We do not have a mono-crystalline bulk of SiO$_2$, and our $480\,$nm laser impinges directly on the surface. Both effects might interfere with the compensation effect due to a small number of surface electrons observed in reference \cite{Sedlacek:2015tg}. 

We do see an immediate improvement after exposing the surface and vacuum cell to UV light. 
The temporal dynamics of that effect suggest that 
rather than directly removing ad-atoms from the surface, the UV light creates
a more uniform layer of Rb atoms by transferring atoms from the vacuum windows to the chip surface. 
Such a uniform layer was used in reference \cite{HermannAvigliano:2014bc} to lower the stray fields. In our case the increase in Rb background pressure could increases the adsorbate coverage on the surface \cite{Chan:2014kb}. This is also consistent with the observation that the stray field increases again with decreasing background pressure and number of atoms in the magnetic trap.

Finally, the $480\,$nm laser clearly lowers the local electric field around its contact area with the surface. 
This suggests that, in contrast to the UV light, we desorb ad-atoms, either by LIAD or by a local increase in temperature that was found to lower the stray fields \cite{Obrecht:2007br,Chan:2014kb,Sedlacek:2015tg}. It is important to note that the lowering of the stray fields is only visible if the surface is exposed to the $480\,$nm laser immediately before the excitation pulse. Most likely Rb atoms produced in the experimental cycle replenish the desorbed ad-atoms.   
\section{Conclusions}
We have found that a coating of SiO$_2$ on our Au surface significantly increases the electric stray field compared to a plain Au surface. We can reduce the stray field by using UV-light, but at the expense of the number of atoms in the magnetic trap. It may thus be possible to strike a balance between these two effects and find an optimum working point. Whereas reference \cite{Sedlacek:2015tg} shows that a significant decrease of the stray field is possible for SiO$_2$ as a bulk material, we could not achieve this improvement and it is unclear if this is possible for micro-traps in close proximity to the surface ($\sim 10\,\mu$m) and in the presence of a high-power $480\,$nm excitation laser. Our results also suggest that heating the surface by a few tens of K is a viable option to significantly reduce the stray fields.
\section*{Acknowledgments}
Our work is financially supported by the Foundation for Fundamental Research on Matter (FOM), which is part of the Netherlands Organisation for Scientific Research (NWO).  
We also acknowledge financial support by the EU H2020 FET Proactive project RySQ (640378).
JN acknowledges financial support by the  Marie Curie program ITN-Coherence (265031). 
We would like to thank J.P. Shaffer for fruitful discussions.
\section*{References}

\providecommand{\newblock}{}

\end{document}